\documentclass[12pt,a4paper]{article}
\usepackage{amssymb}
\usepackage{booktabs,threeparttable}

\newcommand{\ba}{\begin{array}}
\newcommand{\ea}{\end{array}}
\newcommand{\be}{\begin{equation}}
\newcommand{\ee}{\end{equation}}
\newcommand{\bea}{\begin{eqnarray}}
\newcommand{\eea}{\end{eqnarray}}
\newcommand{\bean}{\begin{eqnarray*}}
\newcommand{\eean}{\end{eqnarray*}}
\newcommand{\pa}{\partial}

\newcommand{\ep}{\epsilon}
\newcommand{\no}{\nonumber}

\newcommand{\mL}{\mathcal{L}}
\newcommand{\mM}{\mathcal{M}}
\newcommand{\res}{\mbox{\textrm{res}}}
\newcommand{\tr}{\mbox{\textrm{tr}}}
\newcommand{\lan}{\langle}
\newcommand{\ran}{\rangle}

\begin{document}

\title{Hodograph solutions for the generalized dKP equation}
\author{
Jen-Hsu Chang\thanks{\textit{E-mail address:} jhchang@ndu.edu.tw}\,
and Yu-Tung Chen \\ \\
\textit{Department of Computer Science, National Defense University,} \\
     \textit{ Taoyuan, Taiwan}
}
\date{\today}
\maketitle
\begin{abstract}
We investigate the integrable $(2+1)$-dimensional generalized
dispersionless KP (GdKP) equation (or Manakov-Santinit system)
from the Lax-Sato form.  Several particular three-component
reductions are considered so that the GdKP equation can be reduced
to hydrodynamic systems. Then one can construct infinite exact
solutions of GdKP  by the generalized hodograph method.
\end{abstract}


%%%%%%%%%%%%%%%%%%%%%%%%%%%%%%%%%%%%%%%%%%%%%%%%%%%%%%%%%
%
\section{Introduction}
%
%%%%%%%%%%%%%%%%%%%%%%%%%%%%%%%%%%%%%%%%%%%%%%%%%%%%%%%%%
Recently, in \cite{MS06a,MS07,MS08} Manakov and Santini proposed
an inverse scattering transform for the multidimensional
Hamiltonian vector fields, and analyzed the Cauchy problem for the
following system of PDEs in 2+1 dimensions, the so-called
\emph{Manakov-Santini equation} \cite{MS06a,MS07,MS08,CC08}
\begin{eqnarray}
\frac{1}{3}u_{1,xt}
 &=& \frac{1}{4}u_{1,yy}+(u_1u_{1,x})_x+\frac{1}{2}v_{1,x}u_{1,xy}-\frac{1}{2}u_{1,xx}v_{1,y}, \no\\
\frac{1}{3}v_{1,xt}
 &=& \frac{1}{4}v_{1,yy}+u_1v_{1,xx}+\frac{1}{2}v_{1,x}v_{1,xy}-\frac{1}{2}v_{1,xx}v_{1,y},
\label{MSeq}
\end{eqnarray}
where $u_1=u_1(x,y,t)$ and $v_1=v_1(x,y,t)$ are two distinct field variables.
From the second equation, one can see that $u_1$ can be
expressed as differential polynomial of $v_1$. After plugging it into
the first equation, we obtain the nonlinear \emph{fourth order}
(2+1)-dimensional PDEs for $v_1$.
It is noticed that for $v_1=0$ reduction, the system reduces to the famous dKP equation
\begin{equation}
 \frac{1}{3}u_{1,xt} = \frac{1}{4}u_{1,yy}+(u_1u_{1,x})_x,
\label{dKP-eq}
\end{equation}
while $u_1=0$ reduction gives the equation associated with
Einstein-Weyl space \cite{Pavlov03,Dun04,MS02,MS04}
\begin{equation}
 \frac{1}{3}v_{1,xt} = \frac{1}{4}v_{1,yy}+\frac{1}{2}v_{1,x}v_{1,xy} - \frac{1}{2}v_{1,xx}v_{1,y}.
\label{Pavlov}
\end{equation}
When $u_1=v_{1y}/2=\Phi_{yy}/2$ we can obtain the dispersionless limit of the discrete
KP equation \cite{yu00}:
\[
 \frac{1}{3}\Phi_{xt} = \frac{1}{4}\Phi_{yy} + \frac{1}{4}(\Phi_{xy})^2,
\]
or by setting $\Phi_{xy}=\tilde{\Phi}$,
\[
 \frac{1}{3}\tilde{\Phi}_{xt} = \frac{1}{4}\tilde{\Phi}_{yy} + \frac{1}{4}(\tilde{\Phi}^2)_{xy}.
\]
One remarks that we also have the differential reduction
$u_1=v_{1x}$ associated with the Einstein-Weyl space \cite{Dun08}:
\[\frac{1}{3}v_{1,xt}
 =
 \frac{1}{4}v_{1,yy}+\frac{1}{2}v_{1,x}v_{1,xy}-(\frac{1}{2}v_{1,y}-v_{1x})v_{1,xx}. \]
 \indent In \cite{BDM06}, the hierarchy of GdKP is constructed using the Lax-Sato formulation.
 According to these results, one way to
investigate the Manakov-Santini equation is to generalize it to
the GdKP hierarchy, so that its integrability can be established
such as infinite symmetries, finite-dimensional reductions, etc.
The main purpose of the work is to find the exact solutions of
Manakov-Santini equation (\ref{MSeq}) by generalized hodograph
method via some suitable three-dimensional reductions of
hydrodynamic type.

This paper is organized as follows.
%%%%%%%%%%%%%%%%%%%%%%%%%%%
% Section 2
%%%%%%%%%%%%%%%%%%%%%%%%%%%
In section 2, we use the GdKP hierarchy of Lax-Sato
representation, from which we reproduce the Manakov-Santini
equation in 2+1 dimensions.
%%%%%%%%%%%%%%%%%%%%%%%%%%%
% Section 3
%%%%%%%%%%%%%%%%%%%%%%%%%%%
In section 3, relating to the Orlov operator we impose a special
generating function as a starting point of reduction for  the GdKP
hierarchy. We then use the two-dimensional reductions of Lax
operators arising from the ordinary dKP hierarchy to construct
systems in hydrodynamic form.
%%%%%%%%%%%%%%%%%%%%%%%%%%%
% Section 4
%%%%%%%%%%%%%%%%%%%%%%%%%%%
In section 4, we briefly recall the Hodograph method \cite{GK88}
which transforms the hydrodynamic form into the linear one. In
particular, we solve four examples that are described  in section
3 and obtain hodograph solutions of the Manakov-Santini equation
in rational type.
%%%%%%%%%%%%%%%%%%%%%%%%%%%
% Section 5
%%%%%%%%%%%%%%%%%%%%%%%%%%%
Section 5 is devoted to the concluding remarks.

%%%%%%%%%%%%%%%%%%%%%%%%%%%%%%%%%%%%%%%%%%%%%%%%%%%%%%%%%
%
\section{Generalized dKP hierarchy}
%
%%%%%%%%%%%%%%%%%%%%%%%%%%%%%%%%%%%%%%%%%%%%%%%%%%%%%%%%%
The generalized dKP (GdKP or Manakov-Santini) hierarchy was
constructed by the works \cite{BDM06,BDM07} which can be defined
by the Lax-Sato equations
\begin{eqnarray}
\frac{\pa\psi}{\pa t_n}
= A_n\frac{\pa\psi}{\pa x} - B_n\frac{\pa\psi}{\pa p},
\qquad \psi=\left(\ba{c} \mL\\ \mM \ea\right),
\label{Lax-eq}
\end{eqnarray}
or, equivalently, by the generating equation \be
(J_0^{-1}\mathrm{d}\mL\wedge \mathrm{d}\mM)_-=0, \ee where
$A_n\equiv(J_0^{-1}\pa\mL^n/\pa p)_+$,
$B_n\equiv(J_0^{-1}\pa\mL^n/\pa x)_+$ and  the Lax and Orlov
operators $\mL(p), \mM(p)$ are the Laurent series
\begin{eqnarray}
\mL &=& p+\sum_{n=1}^{\infty}u_n(x)p^{-n},
\label{L}\\
\mM &=& \sum_{n=1}^{\infty}nt_n\mL^{n-1}+\sum_{n=1}^{\infty}v_n(x)\mL^{-n}.
\label{M}
\end{eqnarray}
Here $(\cdots)_+$ ($(\cdots)_-$)
denote respectively the projection on the polynomial part (negative powers), and $J_0$ is defined by
the Poisson bracket
\begin{eqnarray*}
J_0 &=& \{\mL,\mM\}
    = \frac{\pa \mL}{\pa p}\frac{\pa \mM}{\pa x}-\frac{\pa \mL}{\pa x}\frac{\pa \mM}{\pa p}
    = \frac{\pa\mL}{\pa p}\left(\frac{\pa\mM}{\pa x}\Bigg|_{\mL\ \mathrm{fixed}}\right), \\
    &=& 1+v_{1x}p^{-1}+(v_{2x}-u_1)p^{-2}+(v_{3x}-2u_2-2v_{1x}u_1)p^{-3}
        + \cdots.
\end{eqnarray*}
Note that in dKP case $J_0=1$. Some of $A_n$ and $B_n$ are given
by
\begin{eqnarray}
A_1 &=& 1, \no\\
A_2 &=& 2p-2v_{1x}, \no\\
A_3 &=& 3p^2-3v_{1x}p+6u_1+3(v_{1x})^2-3v_{2x}, \no\\
A_4 &=& 4p^3-4v_{1x}p^2+\left(12u_1+4(v_{1x})^2-4v_{2x}\right)p \no\\
     && +12u_2-4v_{3x}+8v_{1x}v_{2x}-4(v_{1x})^3-8u_1v_{1x},
\label{A}
\end{eqnarray}
and
\begin{eqnarray}
B_1 &=& 0, \no\\
B_2 &=& 2u_{1x}, \no\\
B_3 &=& 3u_{1x}p - 3u_{1x}v_{1x}+3u_{2x}, \no\\
B_4 &=& 4u_{1x}p^2+(4u_{2x}-4u_{1x}v_{1x})p \no\\
     && +4u_{1x}\left(4u_1+(v_{1x})^2-v_{2x}\right)-4u_{2x}v_{1x}+4u_{3x}.
\label{B}
\end{eqnarray}
From the Lax equation of $\mL$ in (\ref{Lax-eq}), the evolution equations of $u_k$
with respect to $t_2$ and $t_3$ are derived respectively, by
\bean
\pa_{t_2}u_k
&=& 2\bigg(u_{k+1,x}-v_{1,x}u_{k,x}+(k-1)u_{k-1}u_{1,x}\bigg), \\
\pa_{t_3}u_k
&=& 3\bigg(u_{k+2,x}+ku_ku_{1,x}-v_{1,x}u_{k+1,x}
     +(2u_1+(v_{1,x})^2-v_{2,x})u_{k,x} \\
&&\quad  -(k-1)u_{k-1}u_{1,x}v_{1,x} + (k-1)u_{k-1}u_{2,x}\bigg),
\eean where $k\geq 1$. The first few nontrivial flows of $u_1,
u_2$ and $u_3$ are read off as
\begin{eqnarray}
\pa_{t_2}u_1 &=& 2(u_{2,x}-v_{1,x}u_{1,x}),
\label{u1y} \\
\pa_{t_2}u_2 &=& 2(u_{3,x}-v_{1,x}u_{2,x}+u_1u_{1,x}),
\label{u2y} \\
\pa_{t_2}u_3 &=&  2(u_{4,x}-v_{1,x}u_{3,x}+2u_2u_{1,x}), \no \\
\pa_{t_3}u_1 &=&  3(u_{3,x}+3u_1u_{1,x}-v_{1,x}u_{2,x}-v_{2,x}u_{1,x}+(v_{1,x})^2u_{1,x}),
\label{u1t}\\
\pa_{t_3}u_2 &=&  3(u_{4,x}-v_{1,x}u_{3,x}+3u_1u_{2,x}+2u_{1,x}u_2-v_{1,x}u_1u_{1,x}
                  +(v_{1,x})^2u_{2,x}-v_{2,x}u_{2,x}). \no
\end{eqnarray}
Alternatively, it has been shown \cite{CC08} that the Lax equations for $\mL$ in (\ref{Lax-eq})
is equivalent to
\be
 \frac{\pa p(\mL)}{\pa t_n}\Bigg|_{\mL\ \mathrm{fixed}}
 = A_n(p(\mL))\frac{\pa p(\mL)}{\pa x}\Bigg|_{\mL\ \mathrm{fixed}}+B_n(p(\mL)),
\label{p-eq}
\ee
where
\[
p(\mL)=\mL-\sum_{n=1}^{\infty}\tilde{u}_n\mL^{-n}
\]
is the inverse map of $\mL(p)$ with relations
\be
\tilde{u}_1=u_1,\quad \tilde{u}_2=u_2,\quad \tilde{u}_3=u_1^2+u_3,\quad \tilde{u}_4=3u_1u_2+u_4, \ldots
\label{utu}
\ee
Substituting the expression of $p(\mL)$ into (\ref{p-eq}) and using the formula
$\res(\mL^kd\mL)=\delta_{k,-1}$
we can formally obtain the evolution equations for $\tilde{u}_k$:
\[
 \pa_{t_n}\tilde{u}_k
= \res\Bigg(A_n(p(\mL))\sum_{m=1}^{\infty}\tilde{u}_{m,x}\mL^{k-m-1}-B_n(p(\mL))\mL^{k-1}\Bigg),
 \quad k\geq 1.
\]
For the evolution equations of $v_n$ we notice that the Lax equation for $\mM$ in (\ref{Lax-eq})
can be expressed as the following way:
\[
\frac{\pa\mM}{\pa t_n}\Bigg|_{\mL\ \mathrm{fixed}} +
\frac{\pa\mM}{\pa\mL}\frac{\pa\mL}{\pa t_n} =
A_n(p(\mL))\Bigg(\frac{\pa\mM}{\pa x}\Bigg|_{\mL\ \mathrm{fixed}}
     + \frac{\pa\mM}{\pa\mL}\frac{\pa\mL}{\pa x}\Bigg)
    -B_n(p(\mL))\frac{\pa\mM}{\pa\mL}\frac{\pa\mL}{\pa p}.
\]
After eliminating the Lax equation of $\mL$ we have
\be
 \frac{\pa\mM}{\pa t_n}\Bigg|_{\mL\ \mathrm{fixed}}
= A_n(p(\mL))\,\frac{\pa\mM}{\pa x}\Bigg|_{\mL\ \mathrm{fixed}}.
\label{M-eq} \ee Substituting the expression (\ref{M}) into the
above and again, using $\res(\mL^kd\mL)=\delta_{k,-1}$ it follows
that
\[
\pa_{t_n}v_k = \res_\mL\left(A_n(p(\mL)) \Big(\pa_x\mM\Big|_{\mL\
\mathrm{fixed}}\Big)\mL^{k-1}\right), \quad n,k \geq 1.
\]
For example, for $n=2,3$ we have
\bean
\pa_{t_2}v_k &=& 2\Bigg(v_{k+1,x}-v_{1,x}v_{k,x}-\tilde{u}_k-\sum_{m=1}^{k-1}\tilde{u}_{k-m}v_{k,x}\Bigg), \\
\pa_{t_3}v_k &=& 3\Bigg(v_{k+2,x}-v_{1,x}v_{k+1,x}+\tilde{u}_kv_{1,x}-2\tilde{u}_{k+1}
                        +(2u_1+(v_{1,x})^2-v_{2,x})v_{k,x}       \\
&& \quad       - 2\sum_{m=1}^k\tilde{u}_{k-m+1}v_{m,x}
               + \sum_{m=1}^{k-1}\tilde{u}_{k-m}(\tilde{u}_m+v_{1,x}v_{m,x})
               + \sum_{l=2}^{k-1}\sum_{m=1}^{l-1}\tilde{u}_m\tilde{u}_{l-m}v_{k-l,x}\Bigg).
\eean We read off few of them as
\begin{eqnarray}
\pa_{t_2}v_1 &=& 2(v_{2,x}-(v_{1,x})^2-u_1),
\label{v1y} \\
\pa_{t_2}v_2 &=& 2(v_{3,x}-v_{1,x}v_{2,x}-v_{1,x}u_1-u_2),
\label{v2y} \\
\pa_{t_2}v_3 &=& 2(v_{4,x}-v_{1,x}v_{3,x}-v_{1,x}u_2-v_{2,x}u_1-u_1^2-u_3), \no\\
\pa_{t_3}v_1 &=& 3(v_{3,x}+v_{1,x}u_1-2v_{1,x}v_{2,x}+(v_{1,x})^3-2u_2),
\label{v1t} \\
\pa_{t_3}v_2 &=& 3\left(v_{4,x}-(v_{2,x})^2-v_{1,x}(v_{3,x}+u_2)+(v_{1,x})^2(v_{2,x}+u_1)-u_1^2-2u_3\right),\no
\end{eqnarray}
where Eq. (\ref{utu}) has been used.
We remark here that (\ref{M-eq}) can be written as the following form of conservation equations
\be
 \frac{\pa}{\pa t_n}\left(\frac{\pa\mM}{\pa x}\Bigg|_{\mL\ \mathrm{fixed}}\right)
= \frac{\pa}{\pa x}\left(A_n \frac{\pa\mM}{\pa x}\Bigg|_{\mL\
\mathrm{fixed}}\right), \label{con-law} \ee which means that the
coefficients $\{v_{n,x}\}_{n=1}^{\infty}$ are just the infinitely
many conserved quantities of (\ref{Lax-eq}). To get
(2+1)-dimensional equation of the  GdKP hierarchy, we start from
Eq. (\ref{u1t}) and eliminate $u_{3,x}$ and $v_{2,x}$ by Eqs.
(\ref{u2y}) and (\ref{v1y}) respectively to obtain
\be
\frac{1}{3}\pa_{t_3}u_1 =
\frac{1}{2}\pa_{t_2}u_2-\frac{1}{2}u_{1x}\pa_{t_2}v_1+u_1u_{1x}.
\label{u1_t}
\ee
In the similar way, Eq. (\ref{v1t}) together with
(\ref{v1y}), (\ref{v2y}) obtain
\be \frac{1}{3}\pa_{t_3}v_1 =
\frac{1}{2}\pa_{t_2}v_2 + v_{1x}u_1 -
\frac{1}{2}v_{1x}\pa_{t_2}v_1 -u_2. \label{v1_t}
\ee
Now differentiating Eqs.(\ref{u1_t}) and (\ref{v1_t})  respectively
with respect to $x$ and using $u_{2,x}$ and $v_{2,x}$ by
Eqs.(\ref{u1y}) and (\ref{v1y}), we have the Manakov-Santini
equation (\ref{MSeq}). \\
\textbf{Remark:}
 As mentioned before, when
$v_1=0$ the GdKP hierarchy reduces to the dKP hierarchy
\bean && L
= p+\sum_{n=1}^{\infty}u_np^{-n}, \quad
   M = \sum_{n=1}^{\infty}nt_nL^{n-1}+\sum_{n=2}^{\infty}v_nL^{-n}, \\
&& \pa_{t_n}\Psi = \{\Omega_n, \Psi\}, \quad
   \Omega_n\equiv(L^n)_+, \quad
   \Psi\equiv\left(\ba{c}L \\M \ea\right)
\eean which  means the Poisson bracket $\{L, M\}=1$ is canonical.
From this, it is easy to see that those $v_{n,x}$ $(n\geq 2)$ in
$\pa_xM$ (for $L$ fixed) are nothing but the conserved quantities
$H_n$ (up to a scaling constant) generated by the inverse of $L$,
i.e.,
\[
 p=L-\sum_{k=1}^{\infty}H_kL^{-k}.
\]
Consequently, in contrast to the generalized dKP, the conservation
equations in dKP case
\[
 \pa_{t_n}\left(M_x\bigg|_{L\ \rm{fixed}}\right)
= \pa_x\left(\pa_p\Omega_n(p(L))\bigg(M_x\bigg|_{L\
\rm{fixed}}\bigg)\right),
\]
is equivalent to
\[
 \pa_{t_n}p(L) = \pa_x\Omega_n(p(L)).
\]

%%%%%%%%%%%%%%%%%%%%%%%%%%%%%%%%%%%%%%%%%%%%%%%%%%%%%%%%%
%
\section{Three-dimensional reductions}
%
%%%%%%%%%%%%%%%%%%%%%%%%%%%%%%%%%%%%%%%%%%%%%%%%%%%%%%%%%
Reductions of the GdKP is unlike the dKP system. Except for the
Lax reductions of the dKP hierarchy, the complex relations between
$\{v_{n,x}\}_{n=1}^{\infty}$ and $\{u_n\}_{n=1}^{\infty}$ have to
be considered to construct systems of finite-dimensional
reductions. To solve Manakov-Santini equation (\ref{MSeq}), we
only need three-dimensional reductions. From these reductions, one
can reduce  (\ref{MSeq}) to hydrodynamic form and then obtain
infinite exact solutions from the theory of generalized hodograph
theory. \\
\indent To this end, let us introduce the generating function \be
 Q:= \bigg(\pa_x\mM\bigg|_{\mL}\bigg)^{-1}
\label{gen}
\ee
then equation (\ref{con-law}) is equivalent to
\be
 \pa_{t_n}Q = A_n(\pa_xQ) -(\pa_xA_n) Q := \lan A_n, Q\ran.
\label{Q-eq}
\ee
Next we consider the particular reduction of (\ref{Q-eq}) by the solutions
$Q=Q(p(\mL),\textit{\textbf{V}})$,
where $\textit{\textbf{V}}=(V_1,\ldots,V_m)$.
Then taking into account (\ref{p-eq}), equation (\ref{Q-eq}) reduces to
\be
 \pa_{t_n}Q\bigg|_p = A_n\bigg(\pa_xQ\bigg|_p\bigg) - (\pa_xA_n)Q - B_n(\pa_pQ).
\label{Q-eq-2} \ee In what follows, we discuss the simplest case
for $m=1$ with $V_1:=v$. For simplicity, one sets $v_2=0$. Then
we assume the following ansatz
 \be
 Q = \frac{p}{p-v},
\label{Q-red-1}
\ee
where $v=-v_{1,x}$, which can be seen by comparing the both sides of (\ref{gen}) with fixed $\mL$.
Substituting (\ref{Q-red-1}) into (\ref{Q-eq-2}) we derive
\[
 \pa_{t_n}v = A_n(\pa_xv) - (\pa_xA_n)(p-v) + \frac{v}{p}B_n.
\]
By the result, evaluated at the point $p=v$, the root of $Q^{-1}$
yields the evolution equations of $v$
\be
 \pa_{t_n}v = A_n(p=v)\pa_xv + B_n(p=v).
\label{V-eq} \ee With the ansatz (\ref{Q-red-1}) we see that
Eq.(\ref{p-eq}) for $n=2,3$ can be read off as \bean
 \pa_{t_2}p &=& 2(p+v)p_x + 2u_{1,x}, \\
 \pa_{t_3}p &=& 3(p^2+vp+2u_1+v^2)p_x + 3(p+v)u_{1,x} + 3u_{2,x}.
\eean
Then the compatibility condition $\pa_{t_2}\pa_{t_3}p=\pa_{t_3}\pa_{t_2}p$
with independent variables $p^0, p,p_x,pp_x,$ implies
\bea
&& 6vu_{2,xx} + 6vv_xu_{1,x} - 3u_{2,xt_2} - 6(u_{1,x})^2 + \no\\
&& \quad - 12u_1u_{1,xx} - 3v_yu_{1,x} - 3vu_{1,xt_2} + 2u_{1,xt_3} = 0,
\label{pyt-1}\\
&& 2u_{2,xx} + 2(vu_{1,x})_x - u_{1,xt_2}=0,
\label{pyt-2}\\
&& 3u_{2,x} + v_{t_3} + 6vu_{1,x} + 3v^2v_x - 3vv_{t_2} - 6 u_1v_x - 3u_{1,t_2}=0,
\label{pyt-3}\\
&& 2u_{1,x} + 4vv_x - v_{t_2}=0.
\label{pyt-4}
\eea
By equations (\ref{pyt-2}), (\ref{pyt-4}) we obtain
\bea
 \pa_{t_2}u_1 &=& 2vu_{1,x} + 2u_{2,x}, \label{com-u1y}\\
 \pa_{t_2}v   &=& 4vv_x + 2u_{1,x}. \label{com-v1y}
\eea
Substituting into (\ref{pyt-1}), (\ref{pyt-3}) we have also
\bea
 \pa_{t_3}u_1 &=& (3/2)u_{2,y} + 6u_1u_{1,x} + 3v^2u_{1,x}, \label{com-u1t}\\
 \pa_{t_3}v   &=& 9v^2v_x + 3u_{2,x} + 6(vu_1)_x. \label{com-v1t}
\eea It is easy to see that based on the ansatz (\ref{Q-red-1})
eqs.(\ref{com-v1y}), (\ref{com-v1t}) are nothing but the evolution
equations of $v$ given by (\ref{V-eq}) for $n=2,3$ and moreover;
(\ref{com-u1y}), (\ref{com-u1t}) are related to equations
(\ref{u1y}), (\ref{u2y}), (\ref{u1t}). \\
\indent Consequently, the ansatz (\ref{Q-red-1}) is admissible
with the general expansion of Lax operator (\ref{L}) to have
commuting flows, at least, up to $t_3$. Nevertheless, it is
feasible to apply (\ref{Q-red-1}) with some suitable Lax
reductions to construct explicit solutions to the Manakov-Santini
equations in 2+1 dimensions $(t_1=x,t_2,t_3)$.

Below, under the ansatz (\ref{Q-red-1}), we shall use the Lax
reductions arising from the dKP hierarchy such as $n$th-KdV,
Zakharov as well as the waterbag type reductions, i.e., \bean
 \mL^{n+1} &=& p^{n+1} + w_1p^{n-1} + w_2p^{n-2} + \cdots + w_{n-1} + w_n, \\
 \mL &=& p + \sum_{i=1}^{n-1}\frac{w_i}{p-\tilde{w}_{i}}, \\
 \mL &=& p + \sum_{i=1}^n\ep_i\log(p-w_i), \qquad \ep_1+\cdots+\ep_n=0
\eean to construct finite-dimensional hydrodynamical systems in
two and three components. Four examples shall be discussed as
follows.
\begin{itemize}
\item
(n,m)=(1,1)-reduction (associated with the dKdV reduction),
\item
(n,m)=(2,1)-reduction (associated with the dBoussinesq reduction),
\item
(n,m)=(2,1)-reduction (associated with the Zakharov reduction).
\item
(n,m)=(2,1)-reduction (associated with the Waterbag reduction).
\end{itemize}
Notice that they are different from that of ordinary dKdV,
dBoussinesq, Zakharov and waterbag type reductions. In GdKP
system, a new variable $v$ is involved.

%%%%%%%%%%%%%%%%%%%%%%%%%%%%%%%%%%%%%%%%%%%%%%%%%%%%%%%%%
%
\subsection{(1,1)-reduction: dKdV type}\label{Ex1}
%
%%%%%%%%%%%%%%%%%%%%%%%%%%%%%%%%%%%%%%%%%%%%%%%%%%%%%%%%%
In this case $\mL^2=p^2+u$. Comparing to the expression of
(\ref{L}) we have $u_1=u/2, u_3=-u^2/8, u_5=u^3/16, \ldots$ and
$u_n=0$ for $n\in$ even etc. Using the first few $A_n$ and $B_n$
\bea
&& A_1 = 1, \quad  A_2 = 2(p+v), \quad  A_3 = 3(p^2+vp+u+v^2), \label{dKdV-A}\\
&& B_1 = 0, \quad  B_2 = u_x,  \quad    B_3 = \frac{3}{2}(p+v)u_x,
\no \eea and from Eqs.(\ref{Lax-eq}) for $\mL$ and (\ref{V-eq}),
we obtain the first few nontrivial equations \bea \left(\ba{c}u \\
v\ea\right)_y &=&
\left(\ba{cc}2v & 0 \\ 1 & 4v\ea\right)\left(\ba{c}u \\ v\ea\right)_x, \no\\
\left(\ba{c}u \\ v\ea\right)_t &=&
\left(\ba{cc}3u+3v^2 & 0 \\ 3v & 3u+9v^2\ea\right)\left(\ba{c}u \\ v\ea\right)_x, \no\\
\left(\ba{c}u \\ v\ea\right)_{t_4} &=& \left(\ba{cc}6uv+4v^3 & 0
\\ 3u+6v^2 & 12uv+16v^3\ea\right)\left(\ba{c}u \\ v\ea\right)_x,
\label{C-dKdV} \eea where $t_1=x, t_2=y, t_3=t$. They satisfy not
only the compatibility conditions $\pa_y\pa_t u = \pa_t\pa_y u$
and $\pa_y\pa_t v = \pa_t\pa_y v$, but also community of  flows
for whole hierarchy. On the other hand, by the relation
(\ref{gen}), we derive some of the conserved quantities \bean &&
 v_{1,x}=-v, \quad
 v_{3,x}=-\frac{1}{2}uv, \quad
 v_{5,x}=-\frac{3}{8}u^2v, \quad
 v_{7,x}=-\frac{5}{16}u^3v, \quad\ldots \\
&&
 v_{n,x}=0, \quad n\in \mbox{even}.
\eean Indeed, we list some conservation laws as follows:
\begin{eqnarray}
\pa_y(v_{1,x}) &=& -v_y = -\pa_x(u+2v^2),
\label{con-v1y}\\
\pa_t(v_{1,x}) &=& -v_t = -3\pa_x(uv+v^3),
\label{con-v1t}\\
\pa_y(v_{3,x}) &=& -\frac{1}{2}(uv)_y = -\frac{1}{4}\pa_x(u^2+4uv^2), \no\\
\pa_t(v_{3,x}) &=& -\frac{1}{2}(uv)_t = -\frac{3}{2}\pa_x(u^2v+uv^3), \no\\
\pa_y(v_{5,x}) &=& -\frac{3}{8}(u^2v)_y = -\frac{1}{8}\pa_x(u^3+6u^2v^2), \no\\
\pa_t(v_{5,x}) &=& -\frac{3}{8}(u^2v)_t = -\frac{9}{8}\pa_x(u^3v+u^2v^3). \no
\end{eqnarray}
Note that, using the relations $u=2u_1, v=-v_{1,x}$ and the fact
that $v_{1,y}=-(u+2v^2)$, we see that the compatibility
$\pa_t\pa_yp=\pa_y\pa_tp$ with independent variables $p, \pa_xp,
p\pa_xp$ implies (\ref{C-dKdV}) as well as the Manakov-Santini
equation (\ref{MSeq}).

%%%%%%%%%%%%%%%%%%%%%%%%%%%%%%%%%%%%%%%%%%%%%%%%%%%%%%%%%
%
\subsection{(2,1)-reduction: dBoussinesq type}\label{Ex2}
%
%%%%%%%%%%%%%%%%%%%%%%%%%%%%%%%%%%%%%%%%%%%%%%%%%%%%%%%%%
In this case, $\mL^3=p^3+up+w$. Comparing to (\ref{L}) we have
$u_1=u/3, u_2=w/3, u_3=-u^2/9, u_4=-2uw/9$, etc. Some of $A_n$ and
$B_n$ are given by \bean
&& A_1 = 1, \quad  A_2 = 2(p+v), \quad  A_3 = 3(p^2+vp+\frac{2}{3}u+v^2), \\
&& B_1 = 0, \quad  B_2 = \frac{2}{3}u_x,  \quad  B_3 = (p+v)u_x +
w_x.
 \eean Then from Eqs.(\ref{Lax-eq}) for $\mL$ and (\ref{V-eq})
we derive the first few nontrivial equations \bea
 \left(\ba{c} u \\ w \\ v\ea\right)_y &=&
 \left(\ba{ccc} 2v & 2 & 0 \\
                -2u/3 & 2v & 0 \\
                2/3 & 0 & 4v \ea\right)\left(\ba{c} u \\ w \\ v \ea\right)_x, \no\\
 \left(\ba{c} u \\ w \\ v\ea\right)_t &=&
 \left(\ba{ccc} u+3v^2 & 3v & 0 \\
                -uv & u+3v^2 & 0 \\
                2v & 1 & 2u+9v^2 \ea\right)\left(\ba{c} u \\ w \\ v \ea\right)_x, \label{uwv-yt-1}\\
\left(\ba{c} u \\ w \\ v\ea\right)_{t_4} &=&
4 \left(\ba{ccc} 2uv/3+v^3+w & 2u/3+v^2 & 0 \\
                -uv^2/3-2u^2/9 & v^3+2uv/3+w & 0 \no\\
                2u/9+v^2 & 2v/3 & 2uv + w + 4v^3\ea\right)
\left(\ba{c} u \\ w \\ v \ea\right)_x, \eea where $t_1=x, t_2=y,
t_3=t$. They satisfy the compatibility conditions
$\pa_{t_l}\pa_{t_m} u = \pa_{t_m}\pa_{t_l} u$, $\pa_{t_l}\pa_{t_m}
w = \pa_{t_m}\pa_{t_l} w$ and $\pa_{t_l}\pa_{t_m} v =
\pa_{t_m}\pa_{t_l} v$ for $l,m=1,2,3$. They are NOT compatible
with the $t_4$-flow; however, this does not affect the
construction of the exact solutions in 2+1 dimensions: $(x,y,t)$.
Also, the relation (\ref{gen}) yields  \bean &&
 v_{1,x}=-v, \quad
 v_{2,x}=0, \quad
 v_{3,x}=-\frac{1}{3}uv, \quad
 v_{4,x}=-\frac{1}{3}wv, \\
&&
 v_{5,x}=-\frac{1}{9}u^2v, \quad
 v_{6,x}=-\frac{1}{3} u w v,
\eean
etc. In particular, $v_{1,x}$ satisfies
\be
 \pa_y(v_{1,x}) = -2\pa_x(u/3+v^2), \qquad
 \pa_t(v_{1,x}) = -\pa_x(2uv+w+3v^3).
\label{con-law-2}
\ee

%%%%%%%%%%%%%%%%%%%%%%%%%%%%%%%%%%%%%%%%%%%%%%%%%%%%%%%%%
%
\subsection{(2,1)-reductions: Zakharov type}\label{Ex3}
%
%%%%%%%%%%%%%%%%%%%%%%%%%%%%%%%%%%%%%%%%%%%%%%%%%%%%%%%%%
In this case, $\mL = p+u/(p-w)$. Comparing to (\ref{L}) we have
$u_1=u, u_2=uw, u_3=uw^2, u_4=uw^3$, etc. Some of $A_n$ and $B_n$
are given by \bean
&& A_1 = 1, \quad  A_2 = 2(p+v), \quad  A_3 = 3(p^2+vp+2u+v^2), \\
&& B_1 = 0, \quad  B_2 = 2u_x,  \quad  B_3 = 3(p+v+w)u_x + 3uw_x.
\eean Then from Eqs.(\ref{Lax-eq}) for $\mL$ and (\ref{V-eq}) we
derive the first two nontrivial members as \bea
 \left(\ba{c} u \\ w \\ v\ea\right)_y &=&
 \left(\ba{ccc} 2v+2w & 2u & 0 \\
                2 & 2v+2w & 0 \\
                2 & 0 & 4v \ea\right)\left(\ba{c} u \\ w \\ v \ea\right)_x, \no\\
 \left(\ba{c} u \\ w \\ v\ea\right)_t &=&
 \left(\ba{ccc} 9u+3w^2+3wv+3v^2 & 6uw+3uv & 0 \\
                6w+3v & 9u+3w^2+3wv+3v^2 & 0 \\
                3w+6v & 3u & 6u+9v^2 \ea\right)\left(\ba{c} u \\ w \\ v \ea\right)_x, \no\\
\label{uwv-yt-2} \eea where $t_1=x, t_2=y, t_3=t$. They satisfy
the compatibility conditions $\pa_{t_l}\pa_{t_m} u =
\pa_{t_m}\pa_{t_l} u$, $\pa_{t_l}\pa_{t_m} w = \pa_{t_m}\pa_{t_l}
w$ and $\pa_{t_l}\pa_{t_m} v = \pa_{t_m}\pa_{t_l} v$ for
$l,m=1,2,3$. Also, they are NOT compatible with  the $t_4$-flow.

%%%%%%%%%%%%%%%%%%%%%%%%%%%%%%%%%%%%%%%%%%%%%%%%%%%%%%%%%
%
\subsection{(2,1)-reductions: Waterbag type}\label{Ex4}
%
%%%%%%%%%%%%%%%%%%%%%%%%%%%%%%%%%%%%%%%%%%%%%%%%%%%%%%%%%
In this case,
\[
\mL = p+\ep\log\frac{p-u}{p-w} = p+\ep\sum_{n=1}^{\infty}\frac{1}{n}(w^n-u^n)p^{-n},
\]
and some of $A_n$ and $B_n$ are
\bean
&& A_1 = 1, \quad  A_2 = 2(p+v), \quad  A_3 = 3(p^2+vp-2\ep u+2\ep w+v^2), \\
&& B_1 = 0, \quad  B_2 = 2\ep(-u_x+w_x),  \quad  B_3 = 3\ep(-(p+u+v)u_x+(p+w+v)w_x).
\eean
The $t_2$- and $t_3$-flows for $u, w, v$ can be derived by Eqs.(\ref{Lax-eq}) and (\ref{V-eq}) as
\bea
\left(\ba{c} u \\ w \\ v\ea\right)_y &=&
\left(\ba{ccc}
2u+2v-2\ep & 2\ep & 0 \\
-2\ep & 2w+2v+2\ep & 0 \\
-2\ep & 2\ep & 4v
\ea\right)
\left(\ba{c} u \\ w \\ v\ea\right)_x, \no\\
\left(\ba{c} u \\ w \\ v\ea\right)_y &=&
\left(\ba{ccc}
\ba{l}3(u^2+uv+v^2+ \\[-0.1cm] \;\;\;-4\ep u+2\ep w -\ep v)\ea & 3\ep(u+w+v) & 0 \\[0.4cm]
-3\ep(u+w+v) & \ba{l}3(w^2+wv+v^2+ \\[-0.1cm] \;\;\;+4\ep w -2\ep u +\ep v)\ea & 0 \\[0.4cm]
-3\ep u-6\ep v & 3\ep w+6\ep v & -6\ep u+6\ep w+9v^2
\ea\right)
\left(\ba{c} u \\ w \\ v\ea\right)_x, \no\\
\label{uwv-yt-3}
\eea
where $t_1=x, t_2=y, t_3=t$.
They satisfy the compatibility conditions $\pa_{t_l}\pa_{t_m} u =
\pa_{t_m}\pa_{t_l} u$, $\pa_{t_l}\pa_{t_m} w = \pa_{t_m}\pa_{t_l}
w$ and $\pa_{t_l}\pa_{t_m} v = \pa_{t_m}\pa_{t_l} v$ for
$l,m=1,2,3$. Likewise, they are NOT compatible with the
$t_4$-flow.

%%%%%%%%%%%%%%%%%%%%%%%%%%%%%%%%%%%%%%%%%%%%%%%%%%%%%%%%%
%
\section{Hodograph solutions}
%
%%%%%%%%%%%%%%%%%%%%%%%%%%%%%%%%%%%%%%%%%%%%%%%%%%%%%%%%%
Having setup several reductions of the GdKP hierarchy to
quasilinear(hydrodynamic) systems, we want  to find their exact
solutions by using the generalized hodograph method. In
\cite{GK88}, Gibbons and Kodama developed an systematic way to
generalize the hodograph transformation such that it can be used
to solve hydrodynamic system with enough symmetries.  Now  the
generalized hodograph transformation means the interchanging of
the role of the dependent and independent variables:
$(t_1,\ldots,t_N) \longleftrightarrow (u_1,\ldots,u_N)$. In this
section, following \cite{GK88},  we briefly recall the method as
follows.

The quasilinear system is defined by
\be
 \pa_{t_l}u_i = \sum_{j=1}^Na^l_{ij}\,\pa_xu_j, \quad i,l=1,2,\ldots,N.
\label{Quasi-lin}
\ee
where $a^l_{ij}$ are functions of $(u_1,\ldots,u_N)$ and $a^1_{ij}=\delta_{ij}$.
Defining the $(N-1)$-form
\[
 \Psi_l^{(N-1)} = (-1)^{l+1} dt_1\wedge\cdots\wedge \widehat{dt}_l\wedge\cdots\wedge dt_N,
 \qquad l=1,\ldots,N
\]
where the caret means the $l$th term is omitted. The equation
(\ref{Quasi-lin}) can be rewritten as the differential forms \be
 d u_i\wedge\Psi_l^{(N-1)} = \sum_{j=1}^N a^l_{ij}\, du_j\wedge\Psi_1^{(N-1)}.
\label{diff-form} \ee
and then one has
\be
 \Psi_l^{(N-1)} = \sum_{j=1}^N f_j^l\,\Phi_j^{(N-1)}, \qquad l=1,\ldots,N,
\label{G-hodo}
\ee
where $\Phi_j^{(N-1)}$ are the $(N-1)$-form
\[
 \Phi_j^{(N-1)} = (-1)^{j+1} du_1\wedge\cdots\wedge \widehat{du}_j\wedge\cdots\wedge du_N,
\]
and $f_j^l$ are cofactors of the Jacobian $J=\pa(t_1,\ldots,t_N)/\pa(u_1,\ldots,u_N)$
with non-vanishing $J$.
Substituting (\ref{G-hodo}) into (\ref{diff-form}),
equation (\ref{Quasi-lin}) can be reduced to the following nonlinear PDEs,
called hodograph equations for $(t_1,\ldots,t_N)$:
\be
 f_j^l = \sum_{k=1}^N a^l_{jk}f_k^1, \qquad l=2,3,\ldots,N.
\label{Hodo}
\ee

It has been shown \cite{GK88} that the integrability conditions
for the exact forms $\Psi_l^{(N-1)}$, i.e. $d\Psi_l^{(N-1)}=0$,
can determine $f_j^l$ from a linear system of the defining
equations \be
 \sum_{j=1}^N\pa_jf_j^l = \sum_{k=1}^N\sum_{j=1}^N\pa_j(a^l_{jk}f_k^1) =0, \qquad l=1,\ldots,N.
\label{com-f1}
\ee
where $\pa_j:=\pa/\pa u_j$.
Using the fact that
\[
 \sum_{j=1}^N(\pa_jt_n)f_j^l = \delta_{ln}\frac{\pa(t_1,\ldots,t_N)}{\pa(u_1,\ldots,u_N)}
\]
and the hodograph equation (\ref{Hodo}),
\[
 \sum_{k=1}^N\left[\delta_{ln}\pa_kt_1-\sum_{j=1}^N(\pa_jt_n)a^l_{jk}\right]f_k^1 = 0.
\]
From this equation, one obtains the following linear system as
the dual equations of the hodograph equation
\be
 \delta_{ln}\pa_kt_1 - \sum_{j=1}^N(\pa_jt_n)a^l_{jk}
 = \sum_{r=2}^N\phi^l_{nr}\pa_kt_r, \quad l=2,\ldots,N, \quad n,k=1,\ldots,N,
\label{Lin-eq} \ee where $\phi^l_{nr}=\phi^l_{nr}(u_1,\ldots,u_N)$
are functions to be determined by a particular solution with the
lowest scaling weight (see below). We then show that, by the dual
system (\ref{Lin-eq}), we can further construct polynomial
solutions of higher scaling weights without using the hodograph
equation (\ref{Quasi-lin}).

As examples shown below, we study the cases for $(1,1)$- and
$(2,1)$-reductions with matrix elements $a^l_{jk}$ given by
Sections \ref{Ex1}--\ref{Ex4}. The explicit solutions of
Manakov-Santini equation can be solved in classes of rational as
well as polynomial type.

%
%%%%%%%%%%%%%%%
%
   \paragraph{(I) dKdV type.}
%
%%%%%%%%%%%%%%%
%
In $(1,1)$-reduction, system (\ref{C-dKdV}) is of the quasilinear form with the following
$2\times 2$ matrices corresponding to $t_2$- and $t_3$-flows,
\[
 a^2=
\left(\ba{cc}
2v & 0 \\
1 & 4v
\ea\right) ,\qquad
 a^3=
\left(\ba{cc}
3u+3v^2 & 0 \\
3v & 3u+9v^2
\ea\right).
\]
The compatibility conditions (\ref{com-f1}) for $f_1^1=\pa_vy$ and $f_2^1=-\pa_uy$
($\pa_1:=\pa_u, \pa_2:=\pa_v$) are
\be
\left(\ba{cc} \pa_u & \pa_v \\ 2v\pa_u+\pa_v & 4\pa_vv \ea\right)
\left(\ba{c} f_1^1 \\ f_2^1 \ea \right) = 0,
\label{com-f1-1}
\ee
which admit the polynomial type solutions of the following form with
scaling weights $[u]=2$ and $[v]=1$:
\[
f_1^1 = \sum_{2l_1+l_2 = K-1}\alpha_{l_1,l_2}\,u^{l_1}v^{l_2}, \qquad
f_2^1 = \sum_{2l_1+l_2 = K-2}\beta_{l_1,l_2}\,u^{l_1}v^{l_2},
\]
where $\alpha_{l_1,l_2}$ and $\beta_{l_1,l_2}$ are some constants
to be determined and $K=1,2,\ldots$. For instance, for $K=1$ we
have $f_1^1=\alpha, f_2^1=\beta$. Substituting into
(\ref{com-f1-1}), the hodograph equation (\ref{Hodo}) yields
 \be
 y=\alpha v, \quad x=\alpha(u-v^2),
\label{xy-1} \ee where $\alpha$ is  an arbitrary constant. Using
solutions (\ref{xy-1}), we see that the correction terms
$\phi^2_{n2}$ of the dual linear equation (\ref{Lin-eq}) are
determined by $\phi^2_{12}=8v^2$ and $\phi^2_{22}=-6v$. Thus,
equation (\ref{Lin-eq}) provides \bea
        0 &=& \sum_{j=1}^2(\pa_jx)a^2_{jk} + 8v^2\,\pa_ky, \no\\
 \pa_k x  &=& \sum_{j=1}^2(\pa_jy)a^2_{jk} - 6v\,\pa_ky,
\label{Lin-eq-1} \eea which allows a direct way to solve the
higher weight of polynomial solutions. In practice, let us look
for the next solutions, $K=2$.  We have $x=c_1uv+c_2v^3$ and
$y=c_3u+c_4v^2$. Making use of (\ref{Lin-eq-1}), we solve $c_1=0,\
c_3=-3c_2/8,\ c_4=-3c_2/4$, thus \be
 x=c_2v^3 \qquad y=-\frac{3}{8}c_2(u + 2 v^2),
\label{xy-2} \ee where $c_2$ is an arbitrary constant. Similar
procedures can be done for finding other polynomial type
solutions. We list first few of them with fixed scaling constant
in Table 1.
\begin{center}
\begin{threeparttable}
\caption{Polynomial solutions of the dKdV type reduction.}\label{dKdV}
\begin{tabular*}{120mm}[]{ccccc@{}l}
\toprule
$K$  & $f_1^1$  &  $f_2^1$ & $x$  &  $y$   \\
\midrule
  $1$ &  $1$ & $0$ & $u-v^2$ & $v$
   \\
  $2$ & $12v$ & $-3$ & $-8v^3$ & $3u+6v^2$
   \\
  $3$ & $u+3v^2$ & $-v$ & $\frac{1}{2}u^2-uv^2-\frac{3}{2}v^4$ & $uv+v^3$
   \\
  $4$ & $uv+\frac{4}{3}v^3$ & $-\frac{1}{4}u-\frac{1}{2}v^2$ & $-\frac{2}{3}uv^3-\frac{8}{15}v^5$
    & $\frac{1}{8}u^2+\frac{1}{2}uv^2+\frac{1}{3}v^4$
   \\
\bottomrule
\end{tabular*}
\end{threeparttable}
\end{center}
We remark here that the above solutions of $x$ and $y$ can be
treated as the initial value at $t=0$ \cite{KG89,KG90}. In the
present system, the remaining task is finding the exact solutions
in 2+1 dimensions $(x,y,t)$. It was stated  \cite{Kod88a,Kod88b}
that the dependence of the time variable $t$ may be found
according to the relation of $a^2$ and $a^3$, i.e., \be
  a^3 = \frac{3}{4}(a^2)^2-\frac{3}{2}v a^2 + 3(u+v^2)I,
\label{a3-a2}
\ee
where $I$ is the $2\times 2$ identity matrix.
After changing $(x,y,t)\rightarrow (u,v,t)$
with the dependent variables $x=x(u,v,t)$ and $y=y(u,v,t)$,
the hodograph equation corresponding to the $t$-flow in (\ref{C-dKdV}) (in addition to (\ref{Hodo}))
is given by
\[
 \left(\ba{c} \pa(x,y)/\pa(v,t) \\ -\pa(x,y)/\pa(u,t) \ea\right)
= a^3 \left(\ba{c} \pa_vy \\ -\pa_uy \ea\right),
\]
where $\pa(x,y)/\pa(v,t)$ and $\pa(x,y)/\pa(u,t)$ are the Jacobian
of $(x,y)$ with respect to $(v,t)$ and $(u,t)$, respectively. One
can show that, combining the above and (\ref{Hodo}) with the
relation (\ref{a3-a2}), and requiring that $\pa_uy$ and $\pa_vy$
are independent variables, the implicit hodograph equations have
the following form (string equation) \bea
&& x + \Big(3(u+v^2)-\frac{3}{4}\det (a^2)\Big)t = t_1^0, \no\\
&& y + \Big(\frac{3}{4}\tr (a^2)-\frac{3}{2}v\Big)t = t_2^0,
\label{Hodo-dKdV} \eea where $t_1^0=t_1^0(u,v)$ and
$t_2^0=t_2^0(u,v)$ are initial values at $t=0$ and can be found in
Table \ref{dKdV}. Choosing, for example,  $K=1$, equation
(\ref{Hodo-dKdV}) becomes \bea
&& x + 3(u-v^2)t = u-v^2, \no\\
&& y + 3vt = v. \eea
Then we solve the hodograph solution as
\[
 u(x,y,t) = \frac{x-3xt+y^2}{9t^2-6t+1}, \qquad
 v(x,y,t)=\frac{-y}{3t-1}.
\]
One can verify that $u(x,y,t)$ and $v(x,y,t)$ satisfy the $y$- and $t$-flows of (\ref{C-dKdV}).
Now using $u_1=u/2$, $v_{1,x}=-v$ and the conservation equations (\ref{con-v1y}),
(\ref{con-v1t}), i.e., $v_{1,y}=-(u+2v^2),\,v_{1,t}=-3(uv+v^3)$,
 we obtain a set of rational solution to the Manakov-Santini equation
\[
 u_1(x,y,t) = \frac{1}{2}\frac{x-3xt+y^2}{(3t-1)^2}, \qquad
 v_1(x,y,t) = \frac{-y(x-3xt+y^2)}{(3t-1)^2}.
\]

%
%%%%%%%%%%%%%%%
%
   \paragraph{(II) dBoussnisq type.}
%
%%%%%%%%%%%%%%%
%
For the case of $(2,1)$-reduction in Section \ref{Ex2}, we have,
for the $t_2$- and $t_3$-flows,  the quasilinear system
(\ref{uwv-yt-1}) can be characterized by
\[
(a^2)_{3\times 3} =
\left(\ba{ccc}
2v & 2 & 0 \\
-2u/3 & 2v & 0 \\
2/3 & 0 & 4v
\ea\right), \qquad
(a^3)_{3\times 3} =
\left(\ba{ccc}
3v^2+u & 3v & 0 \\
-uv & 3v^2+u & 0 \\
2v & 1 & 9v^2+2u
\ea\right).
\]
The compatibility conditions (\ref{com-f1}) for $f_1^1=(\pa_wy)(\pa_vt)-(\pa_vy)(\pa_wt)$,
$f_2^1=-(\pa_uy)(\pa_vt)+(\pa_vy)(\pa_ut)$ and $f_3^1=(\pa_uy)(\pa_wt)-(\pa_wy)(\pa_ut)$
($\pa_1:=\pa_u, \pa_2:=\pa_w, \pa_3:=\pa_v$) are
\be
\left(\ba{ccc}
\pa_u & \pa_w & \pa_v \\
2v\pa_u-\frac{2}{3}u\pa_w+\frac{2}{3}\pa_v & 2\pa_u+2v\pa_w & 4\pa_vv  \\
\pa_uu+3v^2\pa_u-uv\pa_w+2\pa_vv & 3v\pa_u+(u+3v^2)\pa_w+\pa_v & 2u\pa_v+9\pa_vv^2
\ea\right)
\left(\ba{c} f_1^1 \\ f_2^1 \\ f_3^1 \ea \right) = 0.
\label{com-f1-2}
\ee
The polynomial type solutions admit the following form with scaling weights
$[u]=2, [w]=3$ and $[v]=1$:
\bean
f_1^1 &=& \sum_{2l_1+3l_2+l_3=2K-3}\alpha_{l_1,l_2,l_3}u^{l_1}w^{l_2}v^{l_3}, \\
f_2^1 &=& \sum_{2l_1+3l_2+l_3=2K-2}\beta_{l_1,l_2,l_3}u^{l_1}w^{l_2}v^{l_3}, \\
f_3^1 &=&
\sum_{2l_1+3l_2+l_3=2K-4}\gamma_{l_1,l_2,l_3}u^{l_1}w^{l_2}v^{l_3},
\eean where $\alpha_{l_1,l_2,l_3}, \beta_{l_1,l_2,l_3}$ and
$\gamma_{l_1,l_2,l_3}$ are constants to be determined and
$K=1,2,\ldots$. To begin with, for $K=1$, we have $f_1^1=\alpha,
f_2^1=\beta$ and $f_3^1=\gamma$. Plugging into (\ref{com-f1-2}),
we obtain $\alpha=\gamma=0$. Then by the definitions of $f_k^1$
and noticing that the scaling weights of $u, w$ and $v$,  one has
\[
 y=c_1\,u+c_2\,v^2, \qquad t=c_3\,v,
\]
where $c_1, c_2, c_3$ are constants with $c_1c_3=-\beta$. Now
substituting $y$ and $t$ into the hodograph equation (\ref{Hodo})
for $l=2$, we obtain the expression of $x$, i.e.,
$x=-2c_1(uv-w)+h(v)$, where $h(v)$ is a function of $v$. Then for
$l=3$ one finds that $c_2=-3c_1/2, c_3=2c_1$ and $h(v)$ has to be
a constant. To summarize, we solve the polynomial solutions as \be
 x=-2c_1(uv-w), \qquad y=c_1(u-3v^2/2), \qquad t=2c_1v,
\label{hodo-B1} \ee where $c_1$ is an arbitrary constant.
Similarly, in this case the dual hodograph equation (given by
(\ref{Lin-eq})) \bea
\delta_{2n}\pa_kx &=& \sum_{j=1}^3(\pa_jt_n)a^2_{jk} + \sum_{r=2}^3\phi^2_{nr}\pa_kt_r, \no\\
\delta_{3n}\pa_kx &=& \sum_{j=1}^3(\pa_jt_n)a^3_{jk} +
\sum_{r=2}^3\phi^3_{nr}\pa_kt_r, \quad n,k=1,2,3, \label{Lin-eq-2}
\eea
with unknown functions $\phi^2_{nr}, \phi^3_{nr}$, can be
found by using the simplest solutions (\ref{hodo-B1}), i.e., \bean
&& \phi^2_{12}=4v^2+(8/3)u, \quad  \phi^2_{13}=6v^3+8uv, \quad
   \phi^2_{22}=-2v, \quad  \phi^2_{23}=3v^2-u, \\
&& \phi^2_{32}=-4/3, \quad  \phi^2_{33}=-6v, \\
&& \phi^3_{12}=6v^3+8uv, \quad  \phi^3_{13}=9v^4+21uv^2+2u^2, \quad
   \phi^3_{22}=3v^2-u, \\
&& \phi^3_{23}=18v^3+(3/2)uv, \quad
   \phi^3_{32}=-6v, \quad  \phi^3_{33}=-9v^2-3u.
\eean Hence the equation (\ref{Lin-eq-2}) provides useful formulas
for determining the polynomial type solutions in higher weights.
For example, we derive the next set of polynomial solution for
$K=2$. The weights for $x,y$ and $t$ are now $4,3$ and $2$,
respectively and have the following general expressions \bean
 x &=& c_1u^2 + c_2 uv^2 + c_3 wv + c_4v^4, \\
 y &=& c_5uv + c_6 w + c_7v^3, \\
 t &=& c_8u + c_9v^2.
\eean Then substituting the above into (\ref{Lin-eq-2}), we find
$c_2=c_7=-c_9=3c_1,\ c_3=c_4=c_5=0,\ c_6=-3c_1/2,\ c_8=-c_1$.
Therefore, we have
\[
 x = c_1(u^2+3uv^2), \quad
 y = -\frac{3}{2}c_1(w-2v^3), \quad
 t = -c_1(u+3v^3),
\]
where $c_1$ is an arbitrary constant. To find the
(2+1)-dimensional equations involving $(x,y,t)$ that satisfy
(\ref{uwv-yt-1}), we choose, for example, the expression of
(\ref{hodo-B1}) and obtain the explicit hodograph solution \bean
 u(x,y,t) &=& \frac{1}{c_1}y + \frac{3}{8c_1^2}t^2, \\
 w(x,y,t) &=& \frac{1}{2c_1}x + \frac{1}{2c_1^2}yt + \frac{3}{16c_1^3}t^3, \\
 v(x,y,t) &=& \frac{1}{2c_1}t.
\eean
Finally, by the relations $u_1=u/3$ and $v_{1,x}=-v$ and the conservation laws (\ref{con-law-2})
one can easily solve the solution satisfing the Manakov-Santini equation as
\bean
u_1(x,y,t) &=& \frac{1}{3c_1}y + \frac{1}{8c_1^2}t^2, \\
v_1(x,y,t) &=& -\frac{1}{2c_1}xt - \frac{1}{3c_1}y^2 - \frac{3}{4c_1^2} yt^2 - \frac{15}{64c_1^3}t^4.
\eean

%
%
%%%%%%%%%%%%%%%
%
   \paragraph{(III) Zakharov type.}
%
%%%%%%%%%%%%%%%
In Section \ref{Ex3}, for the case of $(2,1)$-reduction, we have
\bean
(a^2)_{3\times 3} &=&
\left(\ba{ccc}
2v+2w & 2u & 0 \\
2 & 2v+2w & 0 \\
2 & 0 & 4v
\ea\right), \\
(a^3)_{3\times 3} &=&
\left(\ba{ccc}
3v^2+9u+3w^2+3vw & 3vu+6uw & 0 \\
3v+6w & 3v^2+9u+3w^2+3vw & 0 \\
6v+3w & 3u & 9v^2+6u
\ea\right).
\eean
The compatibility conditions (\ref{com-f1}) show that $f_1^1, f_2^1$ and $f_3^1$ have
polynomial type solutions as the following expressions with scaling weights
$[u]=2, [w]=1$ and $[v]=1$:
\bean
f_1^1 &=& \sum_{2l_1+l_2+l_3=2K-1}\alpha_{l_1,l_2,l_3}u^{l_1}w^{l_2}v^{l_3}, \\
f_2^1 &=& \sum_{2l_1+l_2+l_3=2K-2}\beta_{l_1,l_2,l_3}u^{l_1}w^{l_2}v^{l_3}, \\
f_3^1 &=& \sum_{2l_1+l_2+l_3=2K-2}\gamma_{l_1,l_2,l_3}u^{l_1}w^{l_2}v^{l_3},
\qquad K=1,2,\ldots,
\eean
where $\alpha_{l_1,l_2,l_3}, \beta_{l_1,l_2,l_3}$ and $\gamma_{l_1,l_2,l_3}$
are unknown constants.
By the similar method, we list the first two set of polynomial type solutions as follows.
For $K=1$, we have
\bean
 x &=& 12c\,(uw+3uv-w^2v-wv^2), \\
 y &=& -3c\,(4u-w^2-4wv-v^2), \\
 t &=& -4c\,(w+v).
\eean Then for $K=2$, one has
\bean
 x &=& 6c'(-5u^2-4uw^2-12uwv-9uv^2+4w^3v+7w^2v^2+4wv^3), \\
 y &=& 3c'(5uw-2w^3-8w^2v-8wv^2-2v^3), \\
 t &=& 2c'(5u+3w^2+4wv+3v^2),
\eean where $c$ and $c'$ are arbitrary constants. Here we have
used the same dual hodograph equations (\ref{Lin-eq-2}) in the
previous case, with $\phi^l_{nr},\,l=2,3$ being given by \bean
\phi^2_{12} &=& 8 u+4 v^2, \\
\phi^2_{13} &=& 24 u w +48 u v -12 w^2 v -12 w v^2 +6 v^3, \\
\phi^2_{22} &=& -2 v, \\
\phi^2_{23} &=& -9 u +3 w^2 +12 wv +3 v^2 , \\
\phi^2_{32} &=& -4/3, \\
\phi^2_{33} &=& -4 w -6 v, \\
\phi^3_{12} &=& 24 u w +48 u v -12 w^2 v -12 w v^2 +6 v^3, \\
\phi^3_{13} &=& 54 u^2 +54 u w^2 +144 w u v +153 u v^2 -36 w^3v -81 w^2 v^2 -54 w v^3 +9 v^4,   \\
\phi^3_{22} &=& -9 u +3 w^2 +12 w v +3 v^2, \\
\phi^3_{23} &=& -9 u w -(9/2) u v+9 w^3 + (81/2) w^2 v +54 w v^2 +18 v^3, \\
\phi^3_{32} &=& -4 w -6 v, \\
\phi^3_{33} &=& -15 u -9 w^2 -18 w v-18 v^2. \eean
 \indent These quantities can be used to determine the polynomial type
solutions in higher weights.
%%%%%%%%%%%%%%%
%
   \paragraph{(IV) Waterbag type.}
%
%%%%%%%%%%%%%%%
From Section \ref{Ex4}, we have
\bean
(a^2)_{3\times 3} &=&
\left(\ba{ccc}
2u+2v-2\ep & 2\ep & 0 \\
-2\ep & 2w+2v+2\ep & 0 \\
-2\ep & 2\ep & 4v
\ea\right), \\
(a^3)_{3\times 3} &=&
\left(\ba{ccc}
\ba{l}3(u^2+uv+v^2+ \\[-0.1cm] \;\;\;-4\ep u+2\ep w -\ep v)\ea & 3\ep(u+w+v) & 0 \\[0.4cm]
-3\ep(u+w+v) & \ba{l}3(w^2+wv+v^2+ \\[-0.1cm] \;\;\;+4\ep w -2\ep u +\ep v)\ea & 0 \\[0.4cm]
-3\ep u-6\ep v & 3\ep w+6\ep v & -6\ep u+6\ep w+9v^2
\ea\right).
\eean
The compatibility conditions (\ref{com-f1}) show that $f_1^1, f_2^1$ and $f_3^1$ have
the following polynomial type solutions with scaling weights
$[u]=[w]=[v]=[\ep]=1$:
\[
f_j^1 = \sum\alpha^j_{l_0,l_1,l_2,l_3}\ep^{l_0}u^{l_1}w^{l_2}v^{l_3}
\quad\mbox{with}\quad l_0+l_1+l_2+l_3=2K-1,
\]
where $K=1,2,\ldots$ and $\alpha^j_{l_0,l_1,l_2,l_3}$ for $j=1,2,3$ are unknown constants.
By the similar calculations, we derive the first two polynomial solutions as follows.
For $K=1$, we have
\bean
 x &=& 6C\,(\ep u^2+2uwv+6\ep uv+uv^2-\ep w^2-6\ep wv+wv^2), \\
 y &=& -3C\,(uw+2uv+4\ep u + 2wv -4w\ep+v^2), \\
 t &=& 2C\,(u+w+2v),
\eean
while for $K=2$
\bean
 x &=& 4C'(4 \ep  u^3+6 w v u^2-15 \ep ^2 u^2+18 v \ep  u^2+3 v^2 u^2+27 v^2 \ep  u+6 v^3 u+15 w v^2 u+\\
   && \qquad +30 \ep ^2w u+6 w^2 v u-18 v \ep  w^2-4 \ep  w^3-27 v^2 \ep  w+6 v^3 w-15 \ep ^2 w^2+3 v^2 w^2), \\
 y &=& -3C'(5\ep u^2+4u^2v+2u^2w+2uw^2+8uv^2+8uwv+4v^3+8wv^2-5\ep w^2+4w^2v), \\
 t &=& 4C'(u^2+w^2+3v^2+uw+2uv+2wv-5\ep u+5\ep w),
\eean where $C$ and $C'$ are arbitrary constants. Here we have
used the  dual hodograph equations (\ref{Lin-eq-2}), with
$\phi^l_{nr},\,l=2,3$ being given by \bean
 \phi^2_{12} &=& 8 \ep w-8 \ep u+4 v^2, \\
 \phi^2_{13} &=& 12 \ep w^2+48 v \ep w-48 v \ep u-6 v^2 w+6 v^3-12 \ep u^2-6 v^2 u-12 v u w, \\
 \phi^2_{22} &=& -2 v, \\
 \phi^2_{23} &=& 6 w v+3 v^2+9 \ep u+3 w u-9 \ep w+6 u v, \\
 \phi^2_{32} &=& -4/3, \\
 \phi^2_{33} &=& -2 u-6 v-2 w, \\
 \phi^3_{12} &=& 12 \ep w^2+48 v \ep w-48 v \ep u-6 v^2 w+6 v^3-12 \ep u^2-6 v^2 u-12 v u w, \\
 \phi^3_{13} &=& 72 v w^2 \ep-63 v^2 w u-18 v w u^2+9 v^4+54 \ep^2 w^2+18 \ep w^3-9 v^2 w^2-18 \ep u^3 +54 \ep^2 u^2 \\
             && \; -9 v^2 u^2-27 v^3 u-27 v^3 w-108 \ep^2 u w-153 \ep u v^2-18 v u w^2 +153 v^2 \ep w-72 \ep u^2 v, \\
 \phi^3_{22} &=& 6 w v+3 v^2+9 \ep u+3 w u-9 \ep w+6 u v, \\
 \phi^3_{23} &=& (9/2) \ep u^2+27 v^2 u -(9/2) \ep w^2 +27 v^2 w +18 v^3 +(9/2) v \ep u +9 v u^2 +(9/2) w u^2 \\
             && \; +9 v w^2 +(9/2) w^2 u +(45/2) v u w -(9/2) v \ep w, \\
 \phi^3_{32} &=& -2 u-6 v-2 w, \\
 \phi^3_{33} &=& -3 u^2-9 u v+15 \ep u-18 v^2-3 w u-9 w v-15 \ep w-3 w^2.
\eean

Likewise, these quantities are useful to determine the polynomial
type solutions in higher weights.

%%%%%%%%%%%%%%%%%%%%%%%%%%%%%%%%%%%%%%%%%%%%%%%%%%%%%%%%%
%
\section{Concluding remarks}
%
%%%%%%%%%%%%%%%%%%%%%%%%%%%%%%%%%%%%%%%%%%%%%%%%%%%%%%%%%
In this paper we have studied the Manakov-Santini equation from
the Lax-sato formulation(GdKP hierarchy). By suitable
three-component reductions, one can reduce the  Manakov-Santini
equation to hydrodynamic (quasi-linear) systems. Using the
compatibility conditions $\pa_{t_l}\pa_{t_m} u =
\pa_{t_m}\pa_{t_l} u$, $\pa_{t_l}\pa_{t_m} w = \pa_{t_m}\pa_{t_l}
w$ and $\pa_{t_l}\pa_{t_m} v = \pa_{t_m}\pa_{t_l} v$ for
$l,m=1,2,3$, one can find infinite exact solutions by the
generalized hodograph method. But in all the three-component
reductions, the commuting flows are only up to $t_3$, NOT to
$t_4$. Therefore, it ie quite natural to pose the following
question:
\begin{itemize}
\item How can we find the hydrodynamic reductions for the
generalized dispersionless KP hierarchy ?
\end{itemize}
This issue will be published elsewhere. \\

%%%%%%%%%%%%%%%%%%%%%%%%%%%%%%%%%%%%%%%%%%%%%%%%%%%%%%%%%
\subsection*{Acknowledgments}
We thank Professor Derchyi Wu for providing the literature \cite{GK88}.
We are also grateful to Maxim Pavlov, Iskander A Taimanov and Jyh-Hao Lee for a number of valuable remarks.
This work is supported in part by the National Science Council of Taiwan
under Grant No. NSC 96-2115-M-606-001-MY2 (JHC) and NSC 97-2811-M-606-001 (YTC).

%%%%%%%%%%%%%%%%%%%%%%%%%%%%%%%%%%%%%%%%%%%%%%%%%%%%%%%%%

\end{document}